 \documentclass[preprints,article,accept,moreauthors,pdftex]{mdpi_noJ} 

\firstpage{1} 
\pubvolume{1}
\issuenum{1}
\articlenumber{0}
\pubyear{2021}
\copyrightyear{2020}
\datereceived{} 
\dateaccepted{} 
\datepublished{} 
\hreflink{https://doi.org/} % If needed use \linebreak
\Title{Signatures of topotactic hydrogen in nickelate superconductors}

\TitleCitation{Signatures of topotactic hydrogen in nickelates}

\Author{ Liang Si$^{1,2,*}$\orcidB{}, Paul Worm$^{2}$\orcidC{}, Karsten Held\,$^{2,*}$\orcidA{}}

\AuthorNames{Liang Si, Paul Worm, Karsten Held}

\AuthorCitation{Liang, S.; Worm, P.; Held, K.}
\address{$^1$ School of Physics, Northwest University, Xi'an 710069, China
  \\$^2$ Institute for Solid State Physics, Vienna University of Technology, 1040 Vienna, Austria \\}

\corres{Correspondence: liang.si@ifp.tuwien.ac.at; held@ifp.tuwien.ac.at}

\abstract{
Superconductivity has entered the nickel age marked by enormous experimental and theoretical efforts. Notwithstanding, synthesizing nickelate superconductors remains extremely challenging, not least due to incomplete oxygen reduction and topotactic hydrogen. Here, we present density-functional theory calculations, identify a phonon mode as a possible indication for topotactic hydrogen and discuss the charge redistribution patterns around oxygen and hydrogen impurities.
}


\begin{document}
%%%%%%%%%%%%%%%%%%%%%%%%%%%%%%%%%%%%%%%%%%
%\setcounter{section}{-1} %% Remove this when starting to work on the template.
%\begin{paracol} % PW: I added this. 

\section{Introduction}
\begin{figure}[t]
  %  \begin{minipage}{.69\textwidth}
\begin{center}
\includegraphics[width=13.8cm]{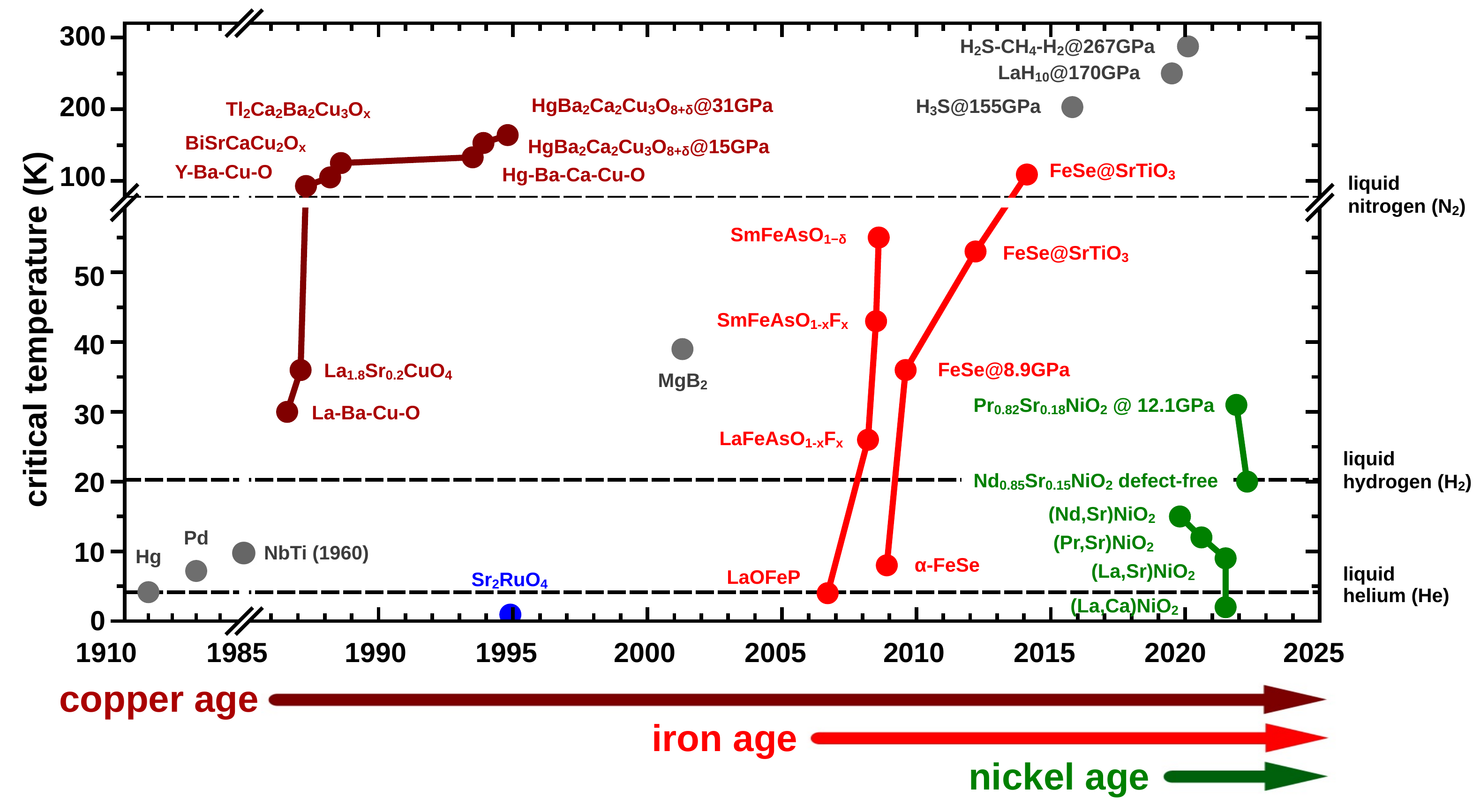}
\end{center}
%\end{minipage}\hfill
%  \begin{minipage}{.3\textwidth}
\caption{Superconducting $T_c$ vs.~year of discovery for selected superconductors.  The discovery of cuprates, iron pnictides and nickelates led to enormous experimental and theoretical activities. Hence one also speaks of the copper, iron and nickel age of superconductivity. \label{fig:ages}  \label{Fig:NiAge}}
%  \end{minipage}
\end{figure}

Computational materials calculations predicted superconductivity in nickelates~\cite{Anisimov1999} and heterostructures thereof~\cite{Chaloupka2008,PhysRevLett.103.016401,Hansmann2010b} since decades, mainly based on  apparent similarly to cuprate superconductors. Three years ago, superconductivity in nickelates was finally discovered in experiment by Li,  Hwang and coworkers~\cite{li2019superconductivity}, breaking the grounds for a new age of superconductivity, the nickel age. It is marked by  an enormous theoretical and experimental activity, including but not restricted to~\cite{li2019superconductivity,zeng2020,Osada2020,Zeng2021,Osada2021,pan2021,Botana2019,Hirofumi2019,Motoaki2019,hu2019twoband,Wu2019,Nomura2019,Zhang2019,Jiang2019,Werner2019,Lechermann2019,Si2019,Lechermann2020,Petocchi2020,Adhikary2020,Subhadeep2020,Karp2020,Kitatani2020,Worm2021c,Geisler2021,Klett2022,LaBolitta2022}. Superconductivity has been found by now, among others,  in 
Nd$_{1-x}$Sr$_x$NiO$_2$ \cite{li2019superconductivity,zeng2020},
Pr$_{1-x}$Sr$_x$NiO$_2$ \cite{Osada2020},
La$_{1-x}$Ca$_x$NiO$_2$ \cite{Zeng2021}, 
La$_{1-x}$Sr$_x$NiO$_2$ \cite{Osada2021},
and most recently in the pentalayer nickelate Nd$_6$Ni$_5$O$_{12}$~\cite{pan2021}.
Fig.~\ref{Fig:NiAge} shows some of the hallmark experimental critical temperatures ($T_c$'s) for the nickelates in comparison with the preceding
copper~\cite{Bednorz1986}
and iron age~\cite{Kamihara2008} of unconventional superconductivity.
Also shown are some other noteworthy superconductors, including the first
superconductor, solid Hg, technologically  relevant NbTi, and hydride superconductors \cite{Drozdov2015}. The last {are} superconducting at room temperature \cite{Snider2020}, albeit only at a pressure of 267GPa exerted in a diamond anvil cell.
All of these compounds are marked in gray in Fig.~\ref{Fig:NiAge} as they are conventional superconductors. That is, the pairing of electrons originates from the electron-phonon coupling, as described in the theory of Bardeen, Cooper, and Schrieffer (BCS) \cite{BCS1957}.

In contrast, cuprates, nickelates, and, to a lesser extent, iron pnictides are strongly correlated electron systems with a large Coulomb interaction between  electrons because of their narrow  transition metal orbitals. Their $T_c$ is too high for BCS theory~\cite{Savrasov1996,Boeri2008}, and the origin of superconductivity in these strongly correlated systems  is still hotly debated.
One prospective mechanism are  antiferromagnetic spin fluctuations \cite{RevModPhys.84.1383,Ohgoe2020,Huang2018,Deng2015,Simkovic2021,Ponsioen2019,Yamase2016,Vilardi2019,Vilardi2020,Sordi2012,Sakai12,Gull2013,Gull2015,Harland2016,Romer2020,Otsuki2014,Kitatani2019} stemming from strong electronic correlations.
%This mechanism has  been firmly established in the  Hubbard model, not least due to largely improved numerical calculations. Here, progress has been brought about  by variational\cite{Ohgoe2020}, determinant \cite{Huang2018} and  diagrammatic quantum Monte Carlo (QMC) simulations\cite{Deng2015,Simkovic2021},
%matrix-product states\cite{Ponsioen2019,Qin2020},  functional renormalization group (fRG)\cite{Yamase2016,Vilardi2019,Vilardi2020},
%cluster\cite{Sordi2012,Sakai12,Gull2013,Gull2015,Harland2016,Romer2020} and diagrammatic extensions\cite{Otsuki2014,Kitatani2019,RMPVertex} of dynamical mean-field theory (DMFT)\cite{Georges1996,Kotliar2004,Held2007}
%\footnote{At,  e.g.\, doping $\delta=1/8$ a stripe phase can be energetically more stable than superconductivity. However, at other dopings, when including a next nearest neighbor hopping $t'$, or a non-local Coulomb interaction $U$, $d$-wave superconductivity prevails\cite{Zheng2017,Qin2020,Huang2018,Ohgoe2020,Ponsioen2019}. The latter also suppresses phase separation, which competes with superconductivity as well.}.
Another mechanism is based on charge density wave fluctuations and
received renewed interest with the discovery of charge density wave ordering 
in cuprates \cite{Comin2016,Wu2011}. Dynamical vertex approximation \cite{Toschi2007,Held2008,Katanin2009,RMPVertex} calculations for nickelates  \cite{Kitatani2020}, that are unbiased with respect to charge and spin fluctuations, found that spin fluctuations dominate and successfully predicted the superconducting dome prior to experiment in Nd$_{1-x}$Sr$_x$NiO$_2$ \cite{Li2020,PhysRevLett.125.147003,lee2022character}.

%With the nickelates we have now a new class of materials that are at the same time very similar to the cuprates but also very different. This rises the hope that this will help us discriminate  the essential from the incidential for unconventional high-$T_c$ superconducticity and thus, eventually, better understanding its microscopic origin. This migth bring us a step closer to the holy grail of solid state physics, a room temperature superconductor at ambient pressuree. Such a material would revolutionize the way we generate, transport and consume electric energy. Already today superconductivity is a billion Euro market. High-$T_c$ superconductors and NbTi are applied for generating stable magnetic fields in magnetic resonance imaging and at the CERN research center. Superconducting cables are already integrated at highest current densities in our electric grid, e.g. in Essen and on Long Island. Further applications include filters for mobile communication, fuses for power plants  and  superconducting quantum interference device (SQUID) magnetometers.

%In this mini review, we first dicuss in Secction~\ref{sec:synthesis} the  difficult synthesis  of nickelate superconductors, which took 20 years  to succeed after the theoretical prediction. Fianlly, Section~\ref{sec:conlcusion} provides for a summary and an outlook.

%\section{Synthesis -- a difficult undertaking}
%\label{sec:synthesis}

\begin{figure}[t]
  %  \begin{minipage}{.69\textwidth}
\begin{center}
\includegraphics[width=13.8cm]{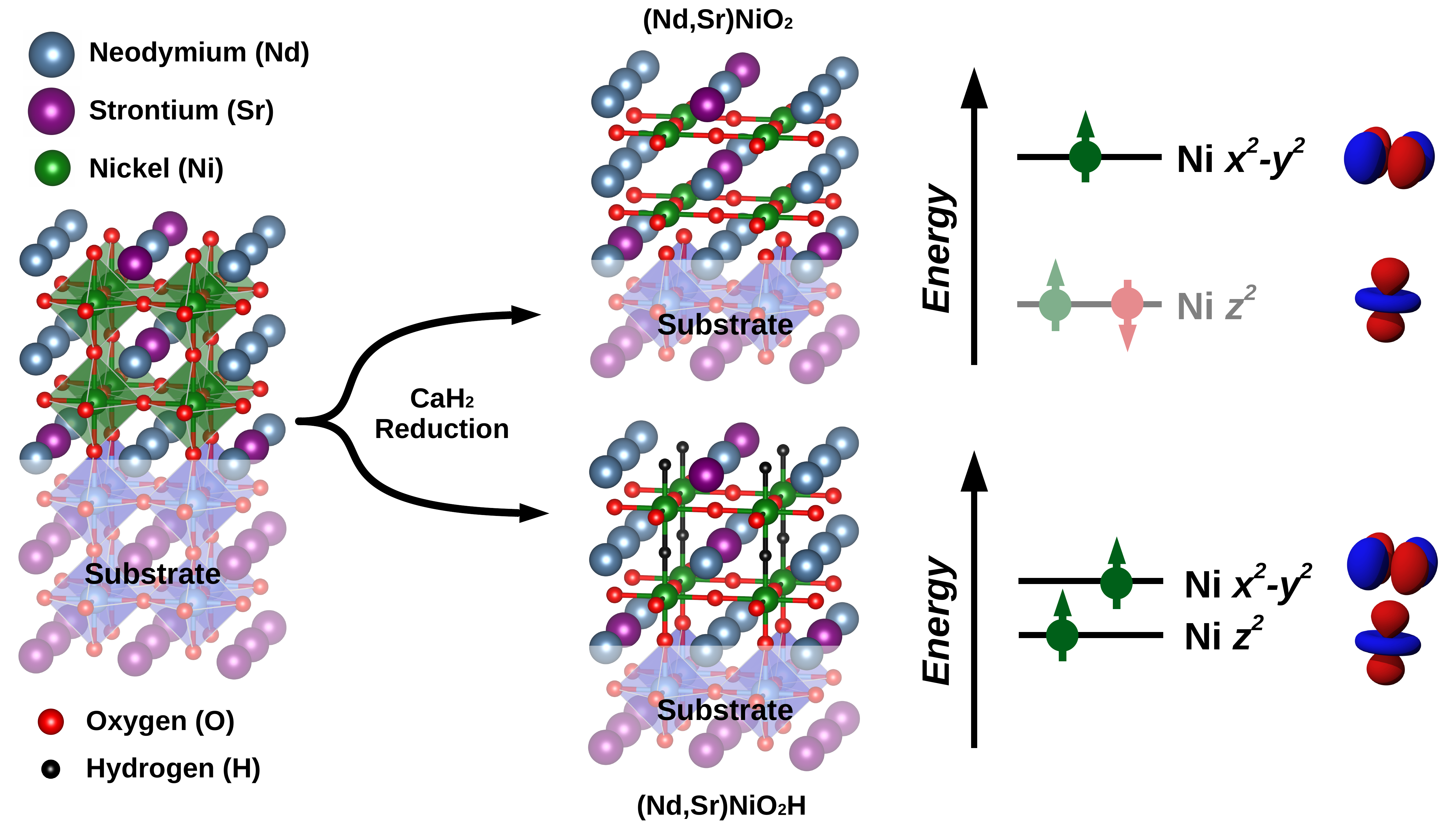}
\end{center}
%\end{minipage}\hfill
%  \begin{minipage}{.3\textwidth}
\caption{For synthesizing superconducting nickelates (1, left) a perovskite film of Nd(La)$_{1-x}$Sr$_x$NiO$_3$  is grown on a SrTiO$_3$ substrate and (2, center) the O atoms between the planes are removed by reduction with CaH$_2$. Besides the pursued nickelate  Nd(La)$_{1-x}$Sr$_x$NiO$_2$ (top center) also excess oxygen or topotactic H may remain in the film, yielding Nd(La)$_{1-x}$Sr$_x$NiO$_2${H} (bottom center). The excess hydrogen results in two holes instead of one hole within the topmost two Ni 3$d$ orbitals (right). Adapted from \cite{Held2022}. \label{Fig:syn} }
%  \end{minipage}
\end{figure}

Why did it take 20 years to synthesize superconducting nickelates that have been so seemingly predicted on a computer?
%But if nickelate superconductors were predicted by theory already 20 years ago, why did it take so long to realize them in experiment ?
To mimic the cuprate Cu 3$d^9$ configuration, as in NdNiO$_2$, nickel has to be in the uncommon oxidation state Ni$^{1+}$ which is rare and prone to oxidize further. Only through a complex two step procedure, Lee, Hwang and coworkers~\cite{Lee2020}  were able to synthesize superconducting nickelates. In a first step, modern 
pulsed laser deposition (PLD) was used to grow a Sr$_x$Nd$_{1-x}$NiO$_3$ film on a SrTiO$_2$ substrate.
This nickelate is still in the %wrong 
3D perovskite phase, see Fig.~\ref{Fig:syn} (left), with one oxygen atom too much and will thus not show superconductivity. Hence, this additional oxygen between the layers needs to be removed in a second step. The reducing agent CaH$_2$ is used to this end, within a quite narrow temperature window~\cite{Lee2020}.  If all goes well, one arrives at the superconducting  Sr$_x$Nd$_{1-x}$NiO$_2$  film (top center). However, this process is prone to incomplete oxidation or to intercalate hydrogen topotactically, i.e., at the  position of the removed oxygen, see Fig.~\ref{Fig:syn} (bottom center). Both of those unwanted outcomes are detrimental for superconductivity.

In~\cite{Si2019,Malyi2022,bernardini2021geometric} it was shown by density functional theory (DFT) calculations that NdNiO$_2$H is indeed energetically favorable to  NdNiO$_2$ + 1/2 H. For the doped system,    on the other hand, Nd$_{0.8}$Sr$_{0.2}$NiO$_2$ without the hydrogen intercalated is energetically favorable. The additional H or likewise an incomplete oxidation to  SrNdNiO$_{2.5}$ alters the physics completely. Additional  H or O$_{0.5}$ will remove an electron from the Ni atoms, resulting in Ni$^{2+}$ instead of Ni$^{1+}$. The formal electronic configuration is hence 3$d^8$ instead of 3$d^9$, or two holes instead of one hole in the Ni $d$-shell. Dynamical mean-field theory (DMFT) calculations \cite{Si2019} evidence that the basic atomic configuration  is the one of  Fig.~\ref{Fig:syn} (lower right). That is, because of Hund's exchange the two holes in  NdNiO$_2$H occupy two different orbitals, 3$d_{x ^2y ^2}$ and 3$d_{3z^2-r^2}$, and form a spin-1. A consequence of this is that DMFT calculations predict NdNiO$_2$H to be a Mott insulator, whereas  NdNiO$_2$ is a strongly correlated metal with a large mass enhancement of about five~\cite{Si2019}.

To the best of our knowledge, such a two-orbital, more 3D electronic structure is unfavorable for high-$T_c$ superconductivity. The two-dimensionality of cuprate and nickelate superconductors helps to suppress long-range antiferromagnetic order, while at the same time retaining strong antiferromagnetic fluctuations that can act as a pairing glue for superconductivity.  In experiment, we cannot expect ideal
NdNiO$_2$, NdNiO$_2$H or    NdNiO$_{2.5}$ films, but most likely some H or additional O will remain in the NdNiO$_2$ film, after the CaH$_2$ reduction. Additional oxygen can be directly evidenced in standard x-ray diffraction analysis after the synthesis step. However, hydrogen, being very light, evades such an x-ray analysis. It has been
evidenced in nickelates only by nuclear magnetic resonance (NMR) experiments \cite{Cui2021} which, contrary to x-ray techniques, are very sensitive to hydrogen.
Ref.~\cite{Puphal2022} suggested hydrogen  in LaNiO$_2$ to be confined at grain boundaries or secondary-phase precipitates.
Given these difficulties, it is maybe not astonishing that it took almost one year before a second research group  \cite{zeng2020}  was able to reproduce superconductivity in nickelates. Despite enormous experimental efforts, only a few groups succeeded hitherto.

%\section{Bandstructure -- similarities and differences to cuprates}
%\label{sec:bandstructure}

%\section{Superconductivity -- }
%\label{sec:superconductivity}

In this paper, we present additional DFT results for topotactic hydrogen and incomplete oxygen reduction in nickelate superconductors:  In Section \ref{Sec:energy} we provide technical information on the DFT calculations. In Section \ref{Sec:energy} we analyze the energy gain to topotactically intercalate hydrogen in LaNiO$_2$ and NdNiO$_2$. In Section \ref{Sec:phonons}, we analyze the phonon spectrum and identify a high-energy mode originating from the Ni-H-Ni bond as a characteristic feature of intercalated hydrogen. In
Section \ref{Sec:charge} we show the changes of the charge distribution caused by topotactic hydrogen. Finally, Section \ref{Sec:conclusion} provides a  summary and outlook.

\section{Method}
\label{Sec:method}

{\em Computational details on $E_b$.} In both our previous theoretical study \cite{Si2019} and this article, the binding energy $E_b$ of hydrogen atoms is computed as:

\begin{equation}
E_{b}=E[ABO_2] +  \mu[H] - E[ABO_2H].
\label{Eq1}
\end{equation}

Here,  $E[ABO_2]$ and $E[ABO_2H]$ are the total energy of infinite-layer $AB$O$_2$ and hydride-oxides $AB$O$_2$H, while  $\mu[H]=E[H_2]/2$ is the chemical potential of H.  Note that $H_2$ is a typical byproduct for the reduction with CaH$_2$ and also emerges when CaH$_2$ is in contact with H$_2$O. Hence it can be expected to be present in the reaction. A positive (negative) $E_b$ indicates the topotactic H process is energetically favorable (unfavorable) to obtain $AB$O$_2$H instead of $AB$O$_2$ and H$_2$/2. 

In the present paper, we go beyond \cite{Si2019} that reported $E_b$ of various $AB$O$_2$ compounds by investigating $E_b$ of La$_{1-x}$Ca$_x$NiO$_2$ systems for many different doping levels. Here, the increasing Ca-doping is achieved by using the virtual crystal approximation (VCA) \cite{PhysRevB.61.7877,PhysRevB.89.165201} from LaNiO$_2$ ($x$=0) to CaNiO$_2$ ($x$=1).
%with doping space of $x$=0.05.
For each Ca concentration, structure relaxation and static total energy calculation is carried out for La$_{1-x}$Ca$_x$NiO$_2$ and La$_{1-x}$Ca$_x$NiO$_2$H  within the tetragonal space group $P$4/mmm.  To this end, we use density-functional theory (DFT) \cite{PhysRev.136.B864,PhysRev.140.A1133} with the \textsc{Vasp} code \cite{PhysRevB.47.558,kresse1996efficiency} and the generalized gradient approximations (GGA) of Perdew, Burke, and Ernzerhof (PBE) \cite{PhysRevLett.77.3865} and PBE revised for solids (PBEsol) \cite{PhysRevLett.100.136406}.
%The in-plane lattice constants are fixed to $a=$\kh{Liang, pls add number}\AA\ of the SrTiO$_3$ substrate as calculated by DFT.
For undoped LaNiO$_2$, the GGA-PBEsol relaxations predict its in-plane lattice constant as 3.890\,\AA~{which}%that
is close to that of {the} STO substrate: 3.905\,\AA.
The computations for La$_{1-x}$Ca$_x$NiO$_2$ and LaCoO$_2$, LaCuO$_2$, SrCoO$_2$ and SrNiO$_2$ are performed without spin-polarization and {a} DFT+$U$ treatment \cite{PhysRevB.44.943}, {as the} inclusion of Coulomb $U$ and spin-polarization only slightly decreases the $E_b$ by $\sim$5\% for LaNiO$_2$ \cite{Held2022} . For  %computations of 
NdNiO$_2$, an inevitably computational issue {are} %is
the localized Nd-4$f$ orbitals. These $f$-orbitals are localized around the atomic core, leading to strong correlations. In non-spin-polarized DFT calculations this generates flat bands near the Fermi level  E$_F$ and leads to unsuccessful convergence. To avoid this, we employed DFT+$U$ [$U_f$(Nd)=7\,eV and $U_d$(Ni)=4.4\,eV] and initialize a $G$-type anti-ferromagnetic ordering for both Nd- and Ni-sublattice in a $\sqrt2$ $\times$ $\sqrt2$ $\times$ $2$ supercell of NdNiO$_2$. For the Nd$_{0.75}$Sr$_{0.25}$NiO$_2$ case, 25\% Sr-doping is achieved by replacing one out of the four Nd atoms by Sr in a $\sqrt2$ $\times$ $\sqrt2$ $\times$ $2$ NdNiO$_2$ supercell.

{\em Computational details on phonons.} The phonon computations for LaNiO$_2$, LaNiO$_2$H, LaNiO$_2$H$_{0.125}$, LaNiO$_{2.125}$ are performed with the frozen phonon method using the \textsc{Phonony} \cite{togo2015first} code interfaced with \textsc{Vasp}. Computations with density functional perturbation  theory (DFPT) method \cite{RevModPhys.73.515} are also carried out for double check. For LaNiO$_2$ and LaNiO$_2$H, the unit cells shown in Fig.~\ref{Fig:phonon}(a,b) are enlarged to a 2$\times$2$\times$2 supercell, while for LaNiO$_2$H$_{0.125}$ and LaNiO$_{2.125}$ the phonon are directly computed with the supercell of Fig.~\ref{Fig:phonon}(c,d).

{\em Computational details on electron density.} The electron density distributions of LaNiO$_2$, LaNiO$_2$H, LaNiO$_2$H$_{0.125}$, and LaNiO$_{2.125}$ are computed using the \textsc{Wien2k} code \cite{blaha2001wien2k} while taking the \textsc{Vasp}-relaxed crystal structure as input. The isosurfaces are plotted from 0.1 (yellow lines) to 2.0 (center of atoms) with spacing 0.1 in units of e/\AA$^2$.

\section{Energetic stability}
\label{Sec:energy}

\begin{figure}[tb]
\begin{center}
\includegraphics[width=13.8cm]{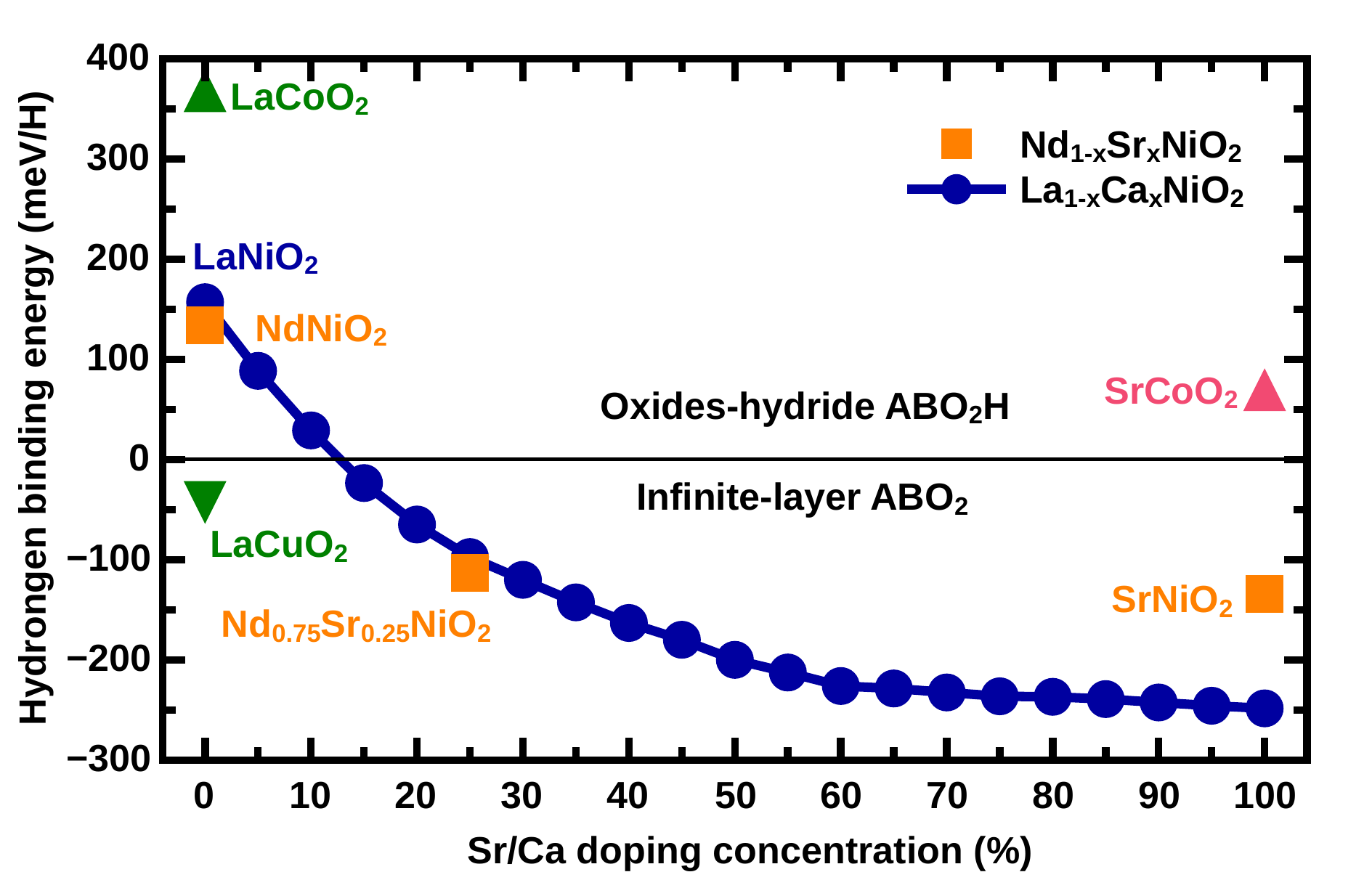}
\end{center}
\caption{Hydrogen binding energy ($E_b$) per hydrogen in two nickelate superconductors, (La,Ca)NiO$_2$ and (Nd,Sr)NiO$_2$; LaCuO$_2$, LaCoO$_2$, and SrCoO$_2$. Slightly above 10\% (Sr,Ca)-doping  infinite layer nickelates are energetically more stable.  Note that the doping changes the filling of the  $B$-3$d$ orbital.
To study the relationship between $E_b$ and the types of $B$-site elements, $E_b$ of several other $AB$O$_2$ compounds is computed: LaCoO$_2$, LaCuO$_2$, SrCoO$_2$ and SrNiO$_2$. Note that this changes the filling of $B$-3$d$ orbital within a large range: e.g. 3$d^8$ for LaCoO$_2$ and 3$d^9$ for LaNiO$_2$. \label{Fig:vca}}
\end{figure}

Fig.~\ref{Fig:vca} shows the results of the hydrogen binding energy $E_b$ for the infinite layer nickelate superconductors Nd$_{1-x}$Sr$_x$NiO$_2$ \cite{li2019superconductivity,Li2020,PhysRevLett.125.147003} and La$_{1-x}$Ca$_x$NiO$_2$ \cite{Zeng2021}. To reveal the evolution of $E_b$ when the $B$-site band filling deviates from their original configurations (3$d^9$ in LaNiO$_2$ when $x$=0 and 3$d^8$ in CaNiO$_2$ when $x$=1), we also show the binding energy of LaCoO$_2$ (3$d^8$), LaCuO$_2$ (3$d^{10}$), SrCoO$_2$ (3$d^7$) and SrNiO$_2$ (3$d^8$).

Let us start with  the case of La$_{1-x}$Ca$_x$NiO$_2$ \cite{Zeng2021}. Here, the unoccupied La-4$f$ orbitals make the computation possible  even without spin-polarization and Coulomb $U$ for La-4$f$, whereas for NdNiO$_2$ this is not practicable due to Nd-4$f$ flat bands near E$_F$. Positive (negative) $E_b$ above (below) the horizontal line in Fig.~\ref{Fig:vca} indicates topotactic H is energetically favorable (unfavorable). When $x$=0, i.e. for  bulk LaNiO$_2$, the system tends to confine H atoms, resulting in oxide-hydride $AB$O$_2$H with $E_b= 157\,$meV/H. As  the concentration of Ca increases,  $E_b$ monotonously decreases, reaching -248\,meV for the end member of the doping series, CaNiO$_2$. The turning point between favorable and unfavorable topotactic H inclusion is around 10\% to 15\% Ca-doping. 
%This is concomitant  
Let us note that  $E_b=0$ roughly agrees with the
onset of superconductivity, which for Ca-doped LaNiO$_2$ emerges for $x$>15\% Ca-doping \cite{Zeng2021}. 

%The computations of $E_b$ in NdNiO$_2$ require a much higher computational effort:
To obtain $E_b$ in  NdNiO$_2$ a much higher computational effort is required:
firstly, the Nd-4$f$ orbitals must be computed with either treating them as core-states or including spin-splitting. Secondly, for the  spin-polarized DFT(+$U$) calculations, an appropriate (anti-)ferromagnetic ordering has to be arranged for both Ni- and Nd-sublattices. In oxide-hydride $AB$O$_2$H compounds, the $\delta$-type bond between Ni and H stabilizes a $G$-type anti-ferromagnetic order by driving the system from a quasi two-dimensional (2D)  system to a three dimensional (3D) one \cite{Si2019}. Given the large computational costs of $E_b$ for Nd$_{1-x}$Sr$_x$NiO$_2$ by using anti-ferromagnetic DFT+$U$ calculations for both Nd-4$f$ ($U\sim$7\,eV) and Ni-3$d$ ($U$=4.4\,eV) orbitals, we merely show here the results of NdNiO$_2$ ($x$=0), Nd$_{0.75}$Sr$_{0.25}$NiO$_2$ ($x$=0.25) and SrNiO$_2$ ($x$=1), which are adopted from \cite{Si2019}. With 25\% Sr-doping, the $E_b$ of NdNiO$_2$ is reduced from 134\,meV to -113\,meV. Please note that $E_b$ of (Nd,Sr)NiO$_2$ is slightly smaller than in (La,Ca)NiO$_2$, at least in the low doping range. This can be explained by shorter lattice constants in NdNiO$_2$, in  agreement with the finding \cite{Si2019} that compressive strain plays an important role at reducing $E_b$.

One can speculate that this suppression of topotactic hydrogen may also play a role when comparing the recently synthesized  (Nd,Sr)NiO$_2$ films on a  (LaAlO$_3$)$_{0.3}$(Sr$_2$TaAlO$_6$)$_{0.7}$ (LSAT) substrate \cite{lee2022character} with the previously employed SrTiO$_3$ (STO) substrate  \cite{Li2020}.
Lee {\em et al.}  \cite{lee2022character} reported cleaner films
without defects  and also a higher superconducting transition temperature  $T_c\sim20\,$K for the LSAT film, as compared to $T_c=15\,$K  and plenty of stacking fault defects for the STO substrate \cite{Li2020}.
As for (La,Ca)NiO$_2$, E$_b=0$ falls in the region of the onset of the
superconductivity for (Sr,Nd)NiO$_2$, which is
$x\sim$10\% Sr-doping in LSAT-strained defect-free films \cite{lee2022character} and $x\sim$12.5\% at SrTiO$_3$-substrate states \cite{Li2020}. Topotactic hydrogen might play a role in  suppressing superconductivity in this doping region.

In Fig.~\ref{Fig:vca}, we further show additional infinite layer compounds LaCoO$_2$, LaCuO$_2$, SrCoO$_2$ and SrNiO$_2$ for comparison. Their $E_b$ is predicted to be 367, -42, 69 and -134\,meV, respectively. Combining the results of LaNiO$_2$ and CaNiO$_2$, we summarize several tendencies on how to predict $E_b$ of $AB$O$_2$: (1) the strongest effect on $E_b$ is changing the $B$-site element. However this seems unpractical for nickelate superconductors as the band filling is strictly restricted to be 3$d^{9-x}$ ($x\sim 0.2$). For both trivalent (La, Nd) and bivalent (Sr, Ca) cations, $E_b$ decreases when the $B$-site cation goes from early to late transition metal elements, e.g. from LaCoO$_2$ (3$d^8$) to LaNiO$_2$ (3$d^9$)  to LaCuO$_2$ (3$d^{10}$). (2) Compressive strains induced by either substrate or external pressure can effectively reduce $E_b$ and we believe that this might be used for growing defect-free films. (3) According to our theoretical calculations, $E_b$ mainly depends on lattice parameters and band filling of the $B$-site 3$d$-orbitals, but much less  on magnetic ordering and Coulomb interaction $U$.

\section{Phonon dispersion}
\label{Sec:phonons}

\begin{figure}[tb]
\begin{center}
\includegraphics[width=10.0cm]{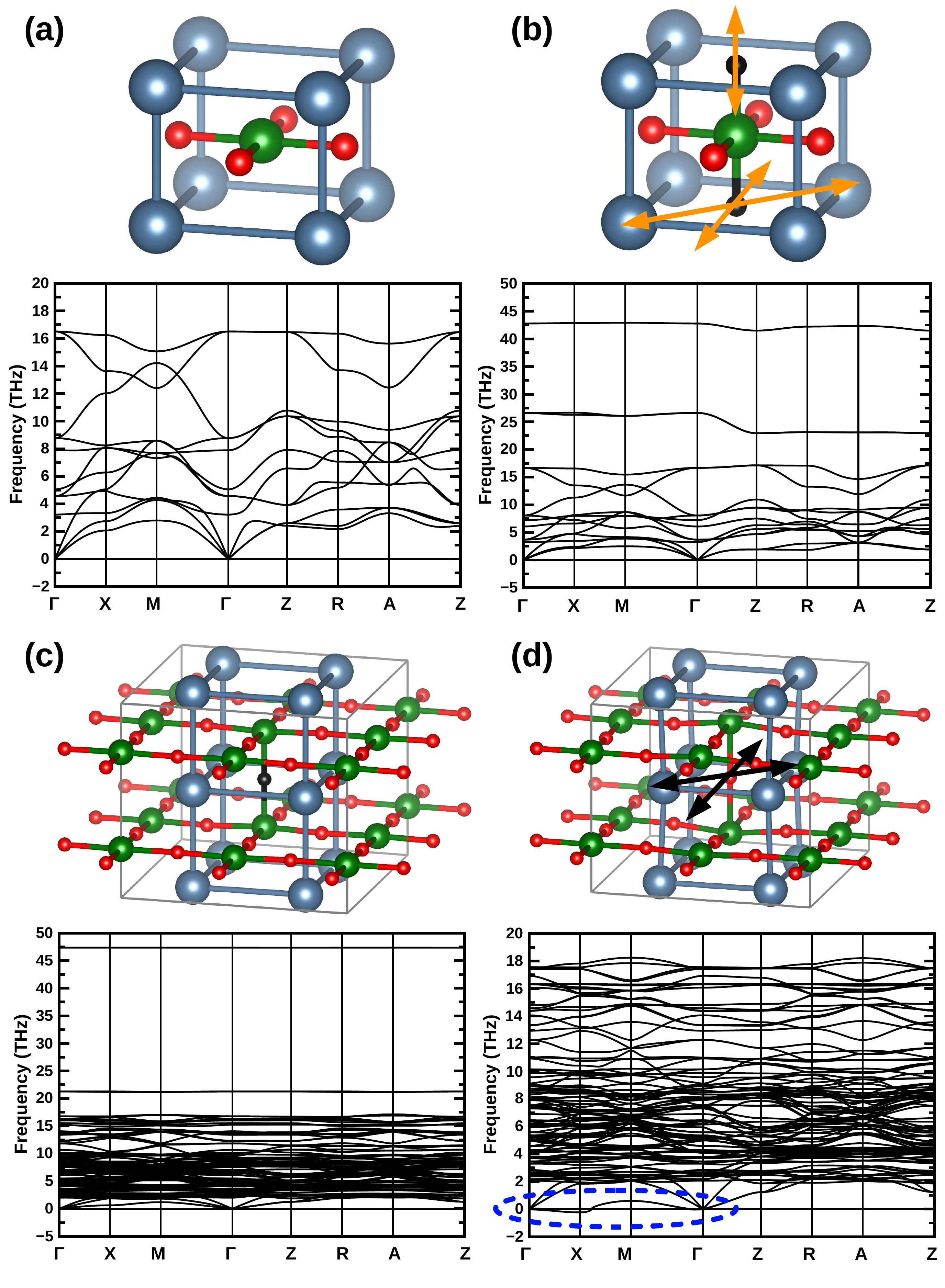}
\end{center}
\caption{Phonon spectra of (a) LaNiO$_2$, (b) LaNiO$_2$H and in a 2$\times$2$\times$2 LaNiO$_2$ supercell doped with a single (c) H and (d) O atom [i.e. LaNiO$_2$H$_{0.125}$ in (c) and LaNiO$_{2.125}$ in (d)]. The orange and black arrows in (b) and (d) represent vibrations of H and O atoms. The blue dashed oval in (d) labels the unstable phonon modes induced by intercalating additional O atoms in LaNiO$_2$.  \label{Fig:phonon}}
\label{Fig4}
\end{figure}

As revealed by previous DFT phonon spectra calculations \cite{Nomura2019}, NdNiO$_2$ is dynamically stable. One of the very fundamental question would be whether topotactic H from over-reacted reduction and/or O from unaccomplished reductive reactions affect the lattice stability. To investigate this point, we perform DFT phonon calculations and analyze the lattice vibration induced by H/O intercalation, as shown in Fig.~\ref{Fig:phonon}.

The phonon spectrum %a
of LaNiO$_2$ [\ref{Fig:phonon}(a)] is essentially the same as in Ref.~\cite{Nomura2019}, all the phonon frequencies are positive, indicating it is dynamically stable. In Fig.~\ref{Fig:phonon}(b), the oxides-hydride LaNiO$_2$H is also predicted to be dynamically stable. Please note that the phonon dispersions between 0 and 20\,THz are basically the same as those in LaNiO$_2$ [Fig.~\ref{Fig:phonon}(a); note the different scale of the $y$-axis]. However, one can see  new, additional vibration modes from the light H-atoms at frequencies of $\sim$27\,THz and $\sim$43\,THz. Among these vibrations, 
the double degenerate mode at lower frequency is generated by an in-plane ($xy$-plane) vibration of the topotactic H atom.  There are two such in-plane vibrations of H atoms, either along the  (100) or (110) direction (and symmetrically related directions), as indicated by the orange arrows in Fig.~\ref{Fig:phonon}(b). The mode located at the higher frequency  $\sim$43\,THz is, on the other hand, formed by an out-of-plane ($z$-direction) vibration and is singly degenerate.

We explain these phonon modes in detail by computing the bonding strength between H-1$s$--Ni-$d_{z^2}$ and H-1$s$--La-$d_{xy}$ orbitals. Our tight-binding calculations yields an electron hopping term of -1.604\,eV between H-1$s$ and Ni-$d_{z^2}$ while it is -1.052\,eV from La-$d_{xy}$ to H-1$s$. That is, the larger H-1$s$--Ni-$d_{z^2}$ overlap leads to a stronger $\delta$-type bonding and, together with the shorter $c$-lattice constant, to a higher phonon energy. Additionally, the shorter $c$-lattice in LaNiO$_2$ should also play a role at forming a stronger H-1$s$--Ni-$d_{z^2}$ bond.

In our previous analysis of the band character for LaNiO$_2$H \cite{Si2019}, the  H-1$s$ bands were mainly located at two energy regions: a very flat band that is mostly from the H-1$s$ itself at $\sim$-7 to -6\,eV, and a hybridized band between H-1$s$ and Ni-$d_{z^2}$ at $\sim$-2\,eV. Together with the higher phonon energy this indicates that  the topotactic H atoms are 
mainly confined by a Ni sub-lattice via bonding and anti-bonding states formed by H-1$s$ and Ni-$d_{z^2}$ orbitals, instead of the La(Nd) sub-lattice.

The complete (full) topotactic inclusion of H, where all vacancies induced by removing oxygen are filled by H, is an ideal limiting case. Under varying experimental conditions, such as chemical reagent, substrate, temperature, and strain, the H-topotactic inclusion may be incomplete, and thus $AB$O$_2$H$_{\delta}$ ($\delta$<1) be energetically favored. Hence, we also compute the phonon {spectrum} at a rather low H-topotactic density: LaNiO$_2$H$_{0.125}$, achieved by including a single H into 2$\times$2$\times$2 LaNiO$_2$ supercells as shown in Fig.~\ref{Fig:phonon}(c). Also such a local H defect, as revealed by the positive frequency at all $q$-vectors in the lower panel of  Fig.~\ref{Fig:phonon}(c), does not destroy the dynamical stability of the LaNiO$_2$ crystal. In fact, the only remarkable qualitative difference between the complete and 12.5\% topotactic H case is the number of phonon bands at 0\,THz to 20\,THz. This is just a consequence of the larger $2\times$2$\times$2 LaNiO$_2$ supercell, with eight times more  phonons. Some quantitative differences can be observed with respect to the energy of the phonon mode: The out-of-plane vibration energy is enhanced from $\sim$43\,THz in LaNiO$_2$H [Fig.~\ref{Fig:phonon}(b)] to $\sim$47\,THz in LaNiO$_2$H$_{0.125}$ [Fig.~\ref{Fig:phonon}(b)], and the in-plane vibration mode frequency is reduced from $\sim$27\,THz in LaNiO$_2$H [Fig.~\ref{Fig:phonon}(b)] to $\sim$21\,THz LaNiO$_2$H$_{0.125}$ [Fig.~\ref{Fig:phonon}(c)].
This is because the H-intercalation shrinks the local $c$-lattice, i.e., the distance between two Ni atoms separated by topotactic H, from 3.383\,\AA~ in  [LaNiO$_2$H: Fig.~\ref{Fig:phonon}(b)]  to 3.327\,\AA~ [LaNiO$_2$H$_{0.125}$: Fig.~\ref{Fig:phonon}(c)]. The bond length between H and La is, on the other hand, slightly increased from 2.767\,\AA~ in [LaNiO$_2$H: Fig.~\ref{Fig:phonon}(b)] to 2.277\,\AA~ [LaNiO$_2$H$_{0.125}$: Fig.~\ref{Fig:phonon}(c)]. This lattice  compression (enlargement) explains the enhancement (reduction) for the out-of-plane (in-plane) phonon frequencies (energies).

These results pave a new way to detect the formation of topotactic H in infinite nickelate superconductors: by measuring the phonon modes. The existence of   localized phonon modes with little dispersion at $\sim$25\,THz and $\sim$45\,THz indicates the presence of topotactic hydrogen, which otherwise would be extremely hard to detect. These frequencies correspond to energies of 103\,meV  and 186\,meV, respectively, beyond the range <80\,meV measured for La$_{1-x}$Sr$_x$NiO$_2$ in \cite{rossi2021broken}. 

Lastly, we further study the case representing an incompleted reduction process: LaNiO$_{2.125}$, achieved by intercalating a single O into a 2$\times$2$\times$2 LaNiO$_2$ supercell [LaNiO$_{2.125}$: Fig.~\ref{Fig:phonon}(d)]. As the same consequence of employing a supercell in phonon computation, the number of phonon bands is multiplied by a factor of 8 in the frequency region between 0\,THz to 20\,THz. One obvious difference between undoped LaNiO$_2$ [Fig.~\ref{Fig:phonon}(a)] and LaNiO$_{2.125}$ [Fig.~\ref{Fig:phonon}(d)] is, that the additional O leads to an unstable phonon mode near $q$=$X$($\pi$,0,0) [blue region in Fig.~\ref{Fig:phonon}(d)]. This phonon mode is formed by an effective vibration of the additional O along the $xy$ plane in the (001) or (110) direction (and symmetrically related directions depending on the exact $q$-vector) of locally cubic coordinate. Such a mode is related to the structural transition from cubic $P$m-3m to a $R$-3c rhombohedral phase as in bulk LaNiO$_3$, with the Ni-O-Ni bond along the $z$-direction deviating from 180$^{\circ}$. Our simulations for other concentrations of additional O atoms (not shown) also indicate that incomplete oxygen reduction reactions  generally result in local instabilities of LaNiO$_{2+\delta}$ with $\delta$>0.

\section{Charge distribution}
\label{Sec:charge}

\begin{figure}[t]
\begin{center}
\includegraphics[width=13.8cm]{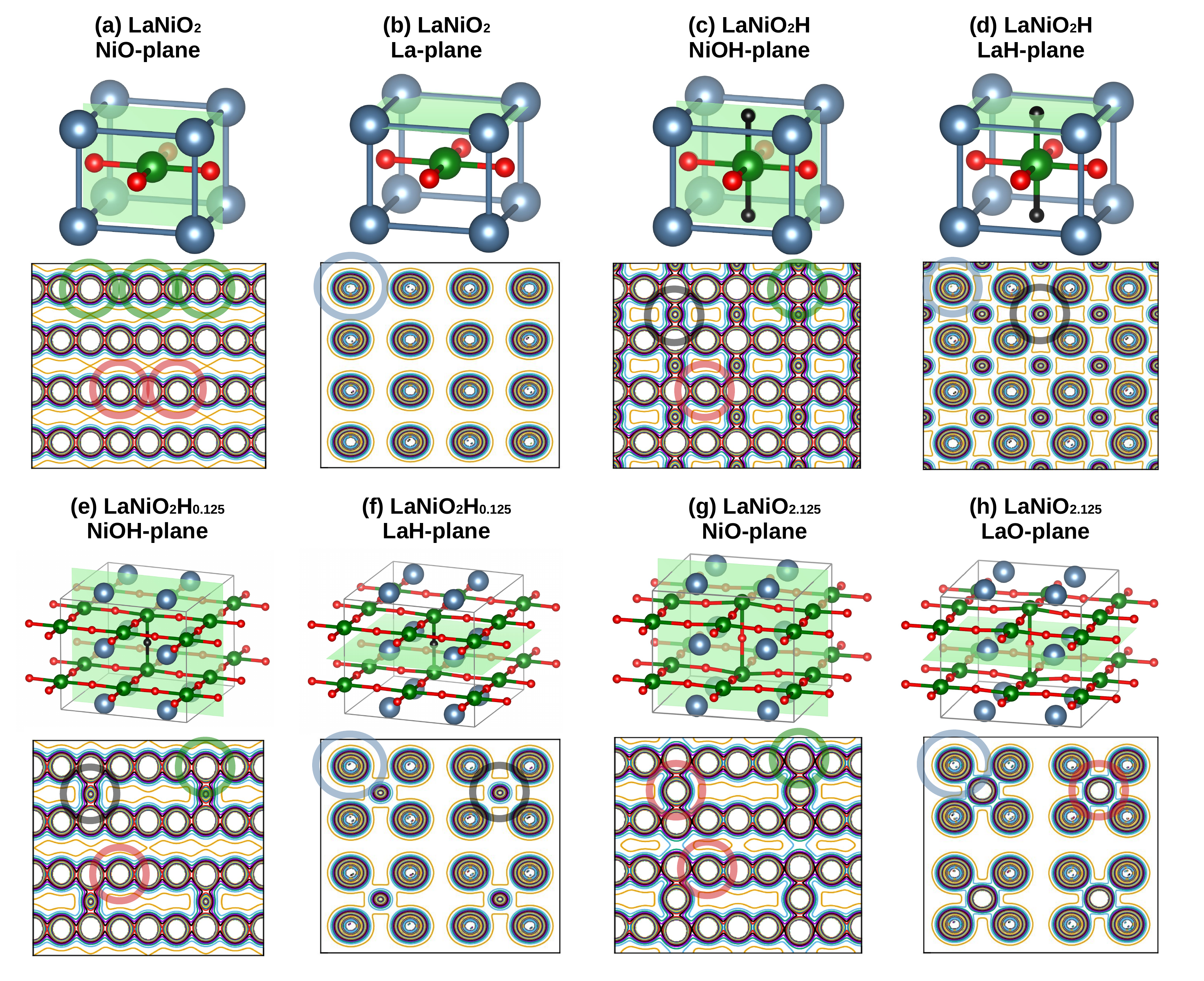}
\end{center}
\caption{DFT calculated valence charge density of (a,b) LaNiO$_2$, (c,d) LaNiO$_2$H, and a LaNiO$_2$ supercell doped with a single (e,f) H and (g,h) O atom. For each compound, the charge density of (020)  and (001) planes are shown in panels (a,c,e,g) and (b,d,f,h),respectively. The La, Ni, O and H atoms are labeled by blue, green, red and black circles, respectively. \label{Fig:charge}}
\end{figure}

In this Section, we perform electron density calculations for LaNiO$_2$, LaNiO$_2$H, LaNiO$_2$H$_{0.125}$ and LaNiO$_{2.125}$ compounds to investigate the bond types resulting from intercalated H and O atoms.
Fig.~\ref{Fig:charge}~(a) and (b) show the electron density of LaNiO$_2$ at the NiO-plane and La-plane (light green planes of the top panels). In Fig.~\ref{Fig:charge}(a), a strong Ni-O bond is observed  while the low electron density between each Ni-O layers reveals a very weak inter-layer coupling, indicating the strong quasi-2D nature of the infinite layer nickelates. In Fig.~\ref{Fig:charge}(b), no  bonds are formed between the La (Nd) atoms. The $A$-site rare-earth elements merely play the role of electron donors.
%\kh{Omit the following, we do not say much for AFM: without forming (anti-)ferromagnetic sub-lattice.}
%Nd$^{3+}$ hosts local moments of 3\,$\mu_B$/Nd from 4$f^3$ configuration.

Fig.~\ref{Fig:charge}~(c) and (d) present the electron density of LaNiO$_2$H along the same  planes. In the NiOH-plane of Fig.~\ref{Fig:charge}(c), the
comparison to  Fig.~\ref{Fig:charge}(c) shows that
intercalated H boosts a 3D picture with an additional $\delta$-type bond formed by Ni-$d_{z^2}$ and H-1$s$ orbitals (black circle). Along the LaH-plane [Fig.~\ref{Fig:charge}(d)], $\delta$-type bonds are formed by the orbital overlap between La-$d_{xy}$ and H-1$s$ orbitals.
For LaNiO$_2$ with partial topotactic H [LaNiO$_2$H$_{0.125}$ in Fig.~\ref{Fig:charge}(e,f)], the additional H atoms play similar roles at the Ni-H and the La-H bonds as in LaNiO$_2$H. The Ni and La atoms without H in-between are similar as in Fig.~\ref{Fig:charge}(a-b) and those with H are akin to  Fig.~\ref{Fig:charge}(c-d). This indicates that the effects induced by topotactic H are indeed very local, i.e., they only affect the the {nearest} %most close 
Ni and La atoms.

In Fig.~\ref{Fig:charge}~(g) and (h), for LaNiO$_{2.125}$, the additional O  increases the local $c$-lattice (Ni-Ni bond length via the additional O) from the LaNiO$_2$ value of 3.338\,\AA~ to 4.018\,\AA\ which is  even larger than the DFT-relaxed value of LaNiO$_3$: 3.80\,\AA. This lattice expansion can be clearly seen in Fig.~\ref{Fig:charge}(g). The large electron density between Ni and O along the $z$-direction indicates the strength of this Ni-O bond in the  $z$-direction is comparable with the ones along $x$/$y$ directions. From Fig.~\ref{Fig:charge}(h), we conclude that similar La-O bonds are formed after intercalating additional O atoms, the La-La distance is shrunken by the additional O atom from 3.889\,\AA\ (LaNiO$_2$) to 3.746\,\AA\ between the La atoms pointing to the additional O. However, from the electron density plot, the La-O bond strength seems not stronger than the La-H bonding in Fig.~\ref{Fig:charge}(c,e). This can be explained by the fact that both O-$p_x$ and -$p_y$ orbitals do not point to orbital lobes of La-$d_{xy}$, leading to a comparable bond strength as the La-H bond in LaNiO$_2$H$_x$.

\section{Conclusion and outlook}
\label{Sec:conclusion}

Our theoretical study demonstrates that the parent compounds of infinite-layer nickelate superconductors, LaNiO$_2$ and NdNiO$_2$, are energetically unstable with respect to topotactic H in the reductive process  from perovskite La(Nd)NiO$_3$ to La(Nd)NiO$_2$. The presence of H, which reshapes the systems from $AB$O$_2$ to the hydride-oxide $AB$O$_2$H, triggers a transition from a quasi-2D strongly correlated single-band ($d_{x^2-y^2}$) metal, to a 2-band ($d_{x^2-y^2}$+$d_{z^2}$) anti-ferromagnetic 3D Mott  insulator. Our predictions \cite{Si2019} have been reproduced by other groups using DFT+$U$ calculations for other similar $AB$O$_2$ systems \cite{Malyi2022,bernardini2021geometric}. The recent experimental observation \cite{krieger2021charge}  of Ni$^{2+}$ (3$d^8$) in nickelates indicates the existence of topotactic H, as do NMR experiments \cite{Cui2021}. 
The presence of H and its consequence of a 3D Mott-insulator is unfavorable for the emergence of superconductivity in nickelates.  However, it is difficult to detect topotactic H in experiment. Three factors contribute to this difficulty: (1)  the small  
radius of H makes it hard to be detected by commonly employed experimental techniques such as x-ray diffraction and scanning transmission electron microscopy (STEM). (2) As revealed by our phonon calculations, the dynamical stability of La(Nd)NiO$_2$ does not rely on the concentration of intercalated H atoms. Hence the same infinite-layer structures should be detected by STEM even in the presence of H. (3) As revealed by electron density distributions, the topotactic H does not break the local crystal structure either (e.g.~bond length and angle); the H atoms  merely affect the most nearby Ni atoms via a Ni-$d_{z^2}$-H-1$s$ $\delta$-bond.
This is  different if we have additional O atoms instead of H: O atoms do not only induce a dynamical instability but also obviously change the local crystal by enlarging the Ni-Ni bond length and angle visibly. Oxygen impurities also lead to unstable phonon modes in LaNiO$_{2+x}$ and thus a major lattice reconstruction. 

The ways to avoid topotactic H revealed by our calculations are: in-plane compressive strains and bivalent cation doping with Sr or Ca. This draws our attention to the recently synthesized  (Nd,Sr)NiO$_2$ films \cite{lee2022character}
, which has been grown  on a (LaAlO$_3$)$_{0.3}$(Sr$_2$TaAlO$_6$)$_{0.7}$ (LSAT) instead of a SrTiO$_3$ (STO) substrate, inducing an additional 0.9\% compressive strain. These new films were shown to be defect-free and  with a considerably larger 
superconducting dome from 10\% to 30\% Sr-doping and a higher maximal $T_c\sim$20\,K  \cite{lee2022character}, compared to 12.5\%-25\% Sr-doping and $T_c\sim$15\,K for nickelate films grown on STO which show many stacking faults \cite{li2019superconductivity,Li2020,PhysRevLett.125.147003}. The compressive strain induced by replacing the STO substrate ($a$=3.905\,\AA) by LSAT ($a$=3.868\,\AA) may tune the positive $E_b$ to negative, thus contributing to suppressing defects and  recovering a single  $d_{x^2-y^2}$-band picture.

%Another opening question is in nickealte samples with partial topotactic-H, whether the H atoms trend to distribute dispersedly or in an ordered way, and whether their distribution affects the magnetic and electronic structures. 

Besides avoiding topotactic H, compressive strain is also predicted as an effectively way to enhance $T_c$. Previous dynamical vertex approximation calculations \cite{Kitatani2020,Held2022} reveal the key to enhance $T_c$ in nickelates is to enhance the bandwidth $W$ and to reduce the ratio of Coulomb interaction $U$ to $W$. Based on this prediction, we have proposed \cite{Kitatani2020,Held2022} three experimental ways to enhance $T_c$ in nickelates: (1) in-plane compressive strain, which can indeed be achieved by using other substrates having a smaller lattice than STO, such as LSAT (3.868\,\AA), LaAlO$_3$ (3.80\,\AA) or SrLaAlO$_4$ (3.75\,\AA). The smaller in-plane lattice shrinks the distance between Ni atoms thus increases their orbital overlap, leading to a larger $W$ and a smaller $U$/$W$. Recent experimental reports have confirmed the validity of this approach by growing (Nd,Sr)NiO$_2$ on LSAT \cite{lee2022character} and Pr$_{0.8}$Sr$_{0.2}$NiO$_2$ on LSAT \cite{ren2021superconductivity}. (2) Applying external pressure on the films plays the  same role as in-plane strain for the, essentially 2D, nickelates. This has been experimentally realized in \cite{wang2021pressure}: under 12.1\,GPa pressure  $T_c$ can be enhanced monotonously to 31\,K without yet showing a saturation. (3) Replacing 3$d$ Ni by 4$d$ Pd. In infinite-layer palladates such as NdPdO$_2$ or LaPdO$_2$ and similar compounds with 2D PdO$_2$ layers and separating  layers between them, the more extended 4$d$ orbitals of Pd are expected to reduce  $U/W$ from  $U/W\sim7$ for nickelates to $U/W\sim6$ for palladates. Further experimental and theoretical research on the electronic and magnetic structure and the superconductive properties of palladates are thus worth to perform.

{\em Acknowledgements} We thank M. Kitatani, J. Tomczak, and Z. Zhong for valuable discussions and the Austrian Science Funds (FWF) for funding through project  P 32044.
%%%%%%%%%%%%%%%%%%%%%%%%%%%%%%%%%%%%%%%%%%
% PW: I commented this out since it seemed to create an error.
\end{paracol}
%%%%%%%%%%%%%%%%%%%%%%%%%%%%%%%%%%%%%%%%%%
% To add notes in main text, please use \endnote{} and un-comment the codes below.
%\begin{adjustwidth}{-5.0cm}{0cm}
%\printendnotes[custom]
%\end{adjustwidth}
%%%%%%%%%%%%%%%%%%%%%%%%%%%%%%%%%%%%%%%%%%
\reftitle{References}

% Please provide either the correct journal abbreviation (e.g. according to the “List of Title Word Abbreviations” http://www.issn.org/services/online-services/access-to-the-ltwa/) or the full name of the journal.
% Citations and References in Supplementary files are permitted provided that they also appear in the reference list here. 

%=====================================
% References, variant A: external bibliography
%=====================================
\externalbibliography{yes}

\end{document}